\documentclass[dvips]{article} 
\usepackage{epsfig}
\usepackage{amssymb} 

\def\aleq{\vcenter{\vbox{\hbox{$\buildrel < \over \sim$}}}} 

\def \sss {\hat{s}} \def \tt {\hat{t}} \def \uu {\hat{u}}

\begin{document}
\textwidth=12.7cm
\textheight=18.4cm
\begin{center}
\section*{HARD EXCLUSIVE BARYON-ANTIBARYON PRODUCTION IN 2-$\gamma$ 
COLLISIONS\footnote{based on a talk given by W. Schweiger at the 
\lq\lq NAPP-2003 Conference\rq\rq, Dubrovnik, Croatia, May 2003} }
\vspace{0.5 cm} CAROLA F. BERGER$^a$ and WOLFGANG SCHWEIGER$^b$\\
\vspace{0.5 cm} $^a${\it I.N.F.N. Sezione di Torino, 
Via P. Giuria 1,
I-10125 Torino, ITALY \\ email: carola.berger@to.infn.it
}\\
\vspace{0.5 cm} $^b${\it Institute of Theoretical Physics, University of
Graz \\A-8010 Graz, Universit\"atsplatz 5, AUSTRIA\\ email:
wolfgang.schweiger@kfunigraz.ac.at}\\
\end{center}
\vspace{0.3 cm} We present a perturbative calculation of baryon pair
production in two-photon collisions, $\gamma \gamma \rightarrow B
\bar{B}$, in which baryons are treated as quark-diquark systems.  Our
approach accounts for constituent mass effects in a systematic way. 
Taking the diquark-model parameters from foregoing studies of other
electron- and photon-induced baryonic reactions, our results 
agree well with the most recent large momentum-transfer data for the
$p\bar{p}$, $\Lambda\bar{\Lambda}$, and $\Sigma^0\bar{\Sigma}^0$
channels.

\noindent PACS numbers: 12.38.Bx, 13.60.Rj 
\newline

\noindent Keywords: perturbative QCD, diquarks, hard hadronic
processes, 2-$\gamma$ reactions\hfill\\

The present study was stimulated by recent experimental efforts of the
OPAL~\cite{OPAL03}, L3~\cite{L302,L303}, and BELLE~\cite{BELLE03}
groups to measure $\gamma \gamma \rightarrow B \bar{B}$ cross sections
at large energies and momentum transfers, where a theoretical
description by means of perturbative QCD is supposed to become
applicable.  It updates and extends previous
work~\cite{ACKS89,KSS91,Kro93} on the perturbative description of
two-photon annihilation into baryon-antibaryon pairs within a
quark-diquark model of baryons.  We have made three improvements
compared to these previous studies:
\begin{itemize}
    \item  all octet baryon channels are considered,

    \item  calculations are performed within the full diquark model 
    with scalar and vector diquarks taken into account,

    \item  constituent-mass effects are included.
\end{itemize}

Our model is a modification of the
usual hard-scattering mechanism~\cite{BL89}, baryons are
described as quark-diquark systems. Diquarks are introduced to parameterize
binding effects between two quarks in a baryon. This approach
allows us to model higher-order
and non-perturbative effects which are undoubtedly present at
currently experimentally accessible momentum transfers.  Within the
hard-scattering picture the hadronic amplitude
$\mathcal{M}_{\{\lambda\} }\left(\sss, \tt\right)$ for $\gamma \gamma
\rightarrow B \bar{B}$ is expressed as a convolution of a
perturbatively calculable hard-scattering amplitude
$\hat{T}_{\{\lambda\}}$, with non-perturbative baryon distribution
amplitudes $\Psi_B$,
\begin{equation}
\mathcal{M}_{\{\lambda\} }\left(\sss, \tt\right) =
%& = &
\int\limits_0^1 d x \int\limits_0^1 d y \, \Psi_B^\dagger
\left(x\right)
\Psi_{\bar{B}}^\dagger \left(y\right) 
%\nonumber \\
%& \times & 
\hat{T}_{\{\lambda\} } \left(x, y; \sss,
\tt\right)\, . \label{HSP}
\end{equation}
Here $x$ and $y$ are the longitudinal 
momentum fractions carried by the quark and
antiquark in the baryon and antibaryon, respectively.  $\sss$ and
$\tt$ denote massless Mandelstam variables, which we will discuss
further below. The subscript $\{ \lambda \}$ denotes all possible
helicity configurations of photon and baryon helicities. 
The hard-scattering amplitude consists of a coherent
superposition of tree graphs that describe the scattering process on
the constituent level.  This means that the two incoming photons are
attached either to the quark or diquark line and quark and diquark are
connected via one-gluon exchange.  Form factors at the gauge-boson
diquark vertices ensure that the correct asymptotic (fixed angle)
scaling behavior of differential cross sections is recovered for large
enough momentum transfers.  The parameterization of the diquark form
factors and of the quark-diquark distribution amplitudes for the
present study has been adopted from previous work~\cite{JKSS93} in
which these parameters have been fitted to electromagnetic nucleon
form factors.  The analytic form of the quark-diquark distribution
amplitudes is flavor dependent such that a heavy strange quark carries
on average more of the baryon momentum than a light up or down
quark.

Before presenting our results we want to comment on a more technical
point, the treatment of constituent masses.  In the standard
hard-scattering approach masses and transverse momenta are neglected
when calculating the hard-scattering amplitude.  We keep the collinear
approximation, but take into account hadron masses.  In this
approximation, each constituent four-momentum is proportional to the
hadron four-momentum.  Likewise, the constituent mass is also
proportional to the hadron mass where the proportionality constant is
the fraction of the hadron momentum which is carried by the
constituent.  A variable constituent mass may look somewhat peculiar,
but one has to keep in mind that the momentum fractions are weighted
by the distribution amplitudes so that, e.g., a quark in a proton has,
on the average, about $1/3$ of the proton mass.  Perturbative
amplitudes are then calculated with these variable constituent masses
and expressed in terms of the hadronic Mandelstam variables $s$, $t$,
and $u$.  As a final step massless Mandelstam variables $\sss$, $\tt$,
$\uu$ are introduced, and the amplitudes are expanded in terms of
$(m_{B}/\sqrt{\sss})$, where $m_{B}$ is the baryon mass.  Only leading
order and next-to-leading order terms are kept in this expansion.  The
leading order terms provide the hadron-helicity conserving amplitudes,
the next-to-leading order terms contribute to helicity amplitudes in
which the hadronic helicity is flipped by one unit.  This 
treatment of mass effects 
has the advantage that no additional mass parameters for the
constituents are introduced, and it preserves gauge invariance with
respect to photon and gluon. Furthermore, it provides also the 
correct crossing relations on the hadronic level 
between the $s-$ and $t-$channel processes,
$\gamma B \rightarrow \gamma B$ and $\gamma \gamma \rightarrow B \bar{B}$,
respectively.  A more detailed account of the above
and other technical questions, including analytical expressions for
the dominant amplitudes and the treatment of time-like diquark form
factors, can be found in Ref.~\cite{BS03}.
\begin{figure}[t!]
\begin{center}
%\vspace{5.0 cm}
\epsfig{file=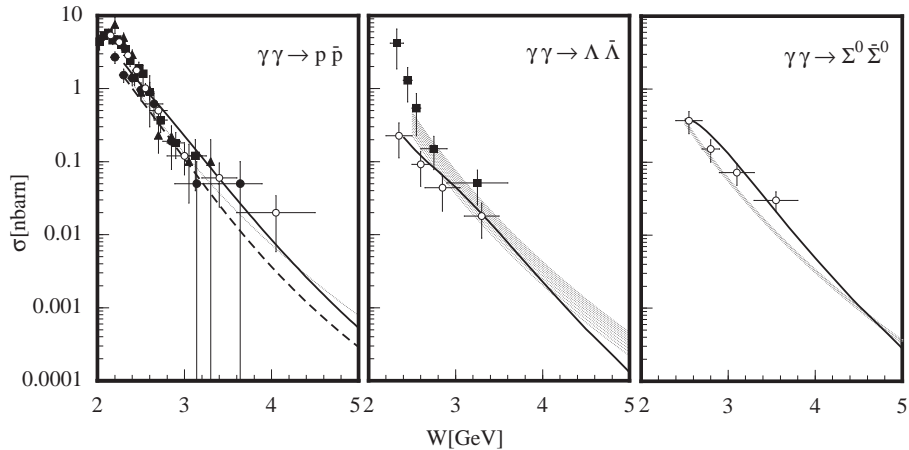,width=12.0cm}
\end{center}
\caption{Integrated $\gamma \gamma \rightarrow p \bar{p}$, $\Lambda
\bar{\Lambda}$, and $\Sigma^0 \bar{\Sigma}^0$ cross sections ($\vert
\cos(\theta)\vert \leq 0.6 $) vs.  $W = \protect\sqrt{s}$. 
Solid (dashed) line: predictions obtained with the standard
parameterization of the diquark model (cf. 
Ref.~\protect{\cite{JKSS93})} with (without) mass corrections.  The
shaded line (band) is the result of the handbag approach
\protect{\cite{DKV03}}. The data are taken from the
CLEO~\protect{\cite{CLEO94, CLEO97}} ($\blacksquare$),
VENUS~\protect{\cite{VENUS97}} ($\blacktriangle$),
OPAL~\protect{\cite{OPAL03}} ($\bullet$), and L3~\protect{\cite{L302,
L303}} ($\circ$) collaborations, respectively.
\label{fig1}}
\end{figure}

Let us now present the results of our calculations.  The energy
dependence of the integrated cross sections ($|\cos(\theta)|\leq 0.6$)
for the $p$, $\Lambda$, and $\Sigma^0$ channels is shown in
Fig.~\ref{fig1}. $\theta$ denotes the center of mass scattering angle.
 The data are taken from the CLEO~\cite{CLEO94,
CLEO97}, VENUS~\cite{VENUS97}, OPAL~\cite{OPAL03}, and L3~\cite{L302,
L303} collaborations.  Preliminary data from BELLE, which agree nicely
with these existing ones, have been reported most recently at the {\em
Photon2003} conference in Frascati~\cite{BELLE03}.  The solid lines,
which correspond to the full diquark-model calculation with mass
effects included, are seen to lie well within the range of data.  The
dashed line in the left plot has been obtained by omitting mass
correction terms, that is, with the two hadronic helicity-conserving
amplitudes only. It can be seen that mass effects are still sizable in
the energy and momentum-transfer range of a few GeV. The contributions
from the (mass dependent) hadronic helicity flip amplitudes are of the
same order of magnitude as those from the hadronic helicity conserving
ones.  At this point it should be mentioned that the predictions for
$\gamma \gamma \rightarrow B \bar{B}$ obtained within the pure quark
hard-scattering approach are less satisfactory~\cite{FZOZ89} and lie
at least one order of magnitude below the data.

There exists an alternative description of 
hard baryon-antibaryon production in two-photon collisions, the 
generalized parton picture, which also achieves good agreement with
experimental data~\cite{DKV03}.  Within this approach,
annihilation is assumed to be dominated by the handbag contribution. 
This contribution is formally power suppressed compared to the 
leading twist one, calculated in the pure quark hard-scattering 
approach. 
Nevertheless, as mentioned at the beginning, the regime where 
leading twist fully dominates seems out of reach for current experiments. 
The authors of~\cite{DKV03} factorize the handbag amplitude 
for $\gamma \gamma \rightarrow B \bar{B}$ 
into a hard $\gamma \gamma \rightarrow q \bar{q}$ amplitude
and into form factors 
encoding the soft
physics of the transition $q \bar{q} \rightarrow B \bar{B}$.
The integrated cross section may then be written as a combination of
these form factors,
\begin{eqnarray}
\sigma(\gamma \gamma \rightarrow B \bar{B}) \vert_{|\cos(\theta)|\leq
0.6} = 181 \hbox{ nbarn GeV}^2 \frac{1}{s} \hbox{\phantom{WWWWWWWWWWW}} 
& & \nonumber\\
\times \left\{ |R_{A}^{B}(s)+R_{P}^{B}(s)|^2 + \frac{s}{4 m_{B}^2}
|R_{P}^{B}(s)|^2+0.134 |R_{V}^{B}(s)|^2 \right\} \, .
\end{eqnarray}
Here, $R_{V}^{B},\, R_{A}^{B}$ and $R_{P}^{B}$ are vector, axial and
pseudoscalar form factors, respectively, 
related to generalized parton distribution functions
 for a baryon $B$. The numerical prefactors are the result of integrating
 the hard amplitude over the angular range $|\cos(\theta)|\leq
0.6$.
The authors of Ref.~\cite{DKV03} argue that $R_{V}^{B}$ is negligible. 
They consider the rest as an effective form factor $\left({R_{\mathrm{
eff}}^{B}}\right)^2=|R_{A}^{B}(s)+R_{P}^{B}(s)|^2 + s/(4 m_{B}^2)
|R_{P}^{B}(s)|^2$ and fix it by means of the integrated $\gamma \gamma
\rightarrow p \bar{p}$ cross section data.  Predictions for the
other hyperon channels are then obtained by means of isospin and 
$U$-spin relations. The down/up ratio
$\rho=F^{d,p}_{i}/F^{u,p}_{i},\,i = V,A,P$ 
of the various proton form factors remains as a parameter, 
where $R_i^p = \sum_q e_q^2 F_i^{q,p}$, with $e_q$ the charge of 
quark $q$. 
The shaded bands in Fig.~\ref{fig1} correspond to $0.25 \leq \rho \leq
0.75$.  With this parametrization the handbag approach provides
results comparable with those from the diquark model, as shown 
in the figure.  

Freund et
al.~\cite{FRSW03} have investigated the time reversed process, $p
\bar{p} \rightarrow \gamma \gamma$, within the generalized parton
picture.  They have tried to model the time-like double distributions
for $p\bar{p}\rightarrow q \bar{q}$ directly and obtain a value of
$0.25 \times 10^{-9}$~fm$^2$ for the integrated cross section
($|\cos(\theta)| \leq 0.7$) at $s=10$~GeV$^2$.  The corresponding
prediction from the diquark model is much larger, namely $0.14 \times
10^{-7}$~fm$^2$ \cite{BS03}.  This, however, is the order of magnitude one would
naively expect from the $\gamma \gamma \rightarrow p \bar{p}$ data and
agrees also with the findings of Diehl and collaborators~\cite{DKV03}. 
These considerations could be interesting in light of the plans for a
$p\bar{p}$ storage ring at GSI in Darmstadt \cite{GSI01}.

The diquark-model provides, of course, also predictions for octet
baryon channels different from $p$, $\Lambda$, or $\Sigma^0$ (see
Ref.~\cite{BS03}).  Those agree roughly with estimates based on
 $SU(3)$ flavor-symmetry relations.  The $\Lambda \bar{\Sigma}^0$
cross section is, for example, small in agreement with the upper
limits quoted by the L3 collaboration.  $\Sigma^+\bar{\Sigma}^-$,
$\Sigma^-\bar{\Sigma}^+$, and $\Xi^-\bar{\Xi}^+$ cross sections are,
on the other hand, of the same order of magnitude as the
$\Lambda\bar{\Lambda}$ cross section, so that there is a chance
for their measurement.  There are indeed efforts from the BELLE
collaboration to determine $\Sigma^+\bar{\Sigma}^-$, and
$\Xi^0\bar{\Xi}^0$ cross sections~\cite{BELLE03}.  A comparison of
different octet baryon channels could be useful to gain information on
the amount of $SU(3)$-symmetry breaking in the baryon distribution
amplitudes.

\begin{figure}[t!]
\begin{center}
%\vspace{5.0 cm}
\epsfig{file=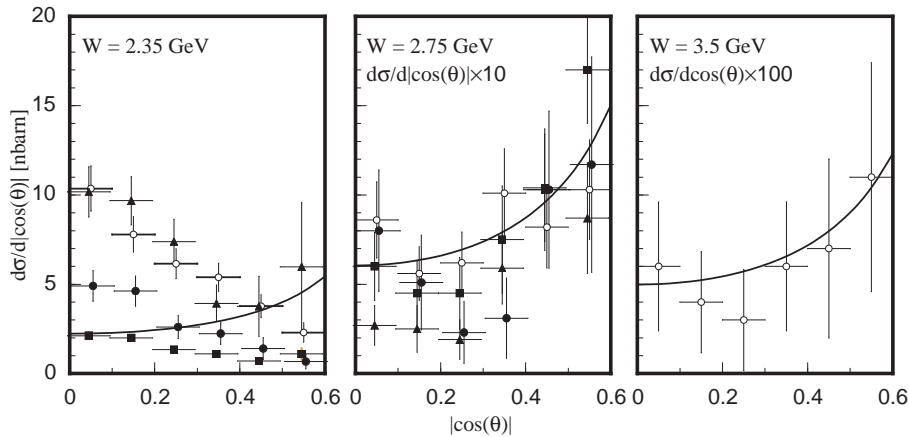,width=12.0cm}
\end{center}
\caption{Differential cross sections $d \sigma(\gamma \gamma \rightarrow p
\bar{p})/d |\cos(\theta)|$) for different values of $W = \sqrt{s}$. Solid line 
and symbols for the data as in Fig.~\ref{fig1}.
\label{fig2}}
\end{figure}
From the integrated cross sections shown in Fig.~\ref{fig1} 
it may seem that the
perturbative QCD approach works already at relatively low energies
($W=\sqrt{s}\approx 2.5$~GeV) and momentum transfers.  Recent
differential cross section data, however, reveal that the perturbative
predictions for $W\,\aleq\,3$~GeV have to be taken with some caution. 
Angular distributions for $\gamma\gamma \rightarrow p\bar{p}$ for
different values of the photon center of mass energy $W$ are shown in
Fig.~\ref{fig2}.  The data show an enhancement in the cross
sections around $\theta=90^\circ$ which decreases with increasing
energy.  Such an enhancement can neither be explained within the
diquark model nor within the handbag approach.  It is a clear signal
for the dominance of low partial waves and is most likely caused by
non-perturbative production mechanisms like, e.g., $p\bar{p}$
production via intermediate-state resonances.  As one would expect,
the perturbative description of the differential cross section becomes
better with increasing $W$.  At $W\approx 3$~GeV the theoretically predicted 
angular distribution is already close to the measured  one. 
There are also attempts by the BELLE collaboration to determine the
angular distributions for the $\Lambda\bar{\Lambda}$ and the
$\Sigma^0\bar{\Sigma}^0$ channels~\cite{BELLE03}.  It will be
interesting to see whether these exhibit a similar behavior as the
$p\bar{p}$ differential cross section.

Finally, we want to mention that there are efforts from the L3
collaboration to determine the $\gamma\gamma \rightarrow
\Delta^{++}\bar{\Delta}^{--}$ cross section.  This is a particularly
interesting process for the diquark-model description, because it
involves only vector diquarks and would thus allow to constrain the
model parameters for the vector diquarks.  Corresponding theoretical
investigations are presently in progress~\cite{BS04}.

\subsubsection*{Acknowledgements}
C.F.B. thanks the Dipartimento di Fisica Teorica, Universit\`a degli Studi di Torino,
Italy, for its hospitality.


\begin{thebibliography}{99}
\bibitem{OPAL03}G. Abbiendi et al. [OPAL Collaboration], 
Eur. Phys. J {\bf C 28} (2003) 45 [arXiv:hep-ex/0209052].
%%CITATION = HEP-EX 0209052;%%
\bibitem{L302}P. Achard et al. [L3 Collaboration],
 Phys. Lett. {\bf B 536} (2002) 24 [arXiv:hep-ex/0204025].
%%CITATION = HEP-EX 0204025;%%
\bibitem{L303}P. Achard et al. [L3 Collaboration], hep-ex/0306017.
%%CITATION = HEP-EX 0306017;%%
\bibitem{BELLE03}Ch.-Ch. Kuo et al., in {\it Proceedings of PHOTON 
2003, Frascati, Italy}, eds. F. Anulli et al., to appear in Nucl. Phys. 
Proc. Suppl..
\bibitem{ACKS89}M.~Anselmino, F.~Caruso, 
P.~Kroll and W.~Schweiger, Int. J. Mod. Phys. {\bf A 4} 
(1989) 5213.
%%CITATION = IMPAE,A4,5213;%%
\bibitem{KSS91}P.~Kroll, M.~Sch\"urmann and W.~Schweiger, 
Int. J. Mod. Phys. {\bf A 6} 
(1991) 4107.
%%CITATION = IMPAE,A6,4107;%%
\bibitem{Kro93}P.~Kroll, T.~Pilsner, M.~Sch\"urmann and W.~Schweiger, 
Phys. Lett. {\bf B 316} (1993) 546 [arXiv:hep-ph/9305251].
%%CITATION = HEP-PH 9305251;%%
\bibitem{BL89}See, for example, S.~J. Brodsky and G.~P. Lepage, in {\it
Perturbative Quantum Chromodynamics}, ed. by A. H. Mueller (World
Scientific, Singapore, 1989).
\bibitem{JKSS93}R.~Jakob, P.~Kroll, M.~Sch\"urmann and W.~Schweiger, 
Z. Phys. {\bf A 347} (1993) 109 [arXiv:hep-ph/9310227].
%%CITATION = HEP-PH 9310227;%%
\bibitem{BS03} C.~F. Berger and W. Schweiger, Eur. Phys. J {\bf C 28} 
(2003) 249 [arXiv:hep-ph/0212066].
%%CITATION = HEP-PH 0212066;%%
\bibitem{CLEO94}M. Artuso et al. [CLEO Collaboration], 
Phys. Rev. {\bf D 50} (1994) 5484.
%%CITATION = PHRVA,D50,5484;%%
\bibitem{CLEO97}S. Anderson et al. [CLEO Collaboration], 
Phys. Rev. {\bf D 56} (1997)
2485 [arXiv:hep-ex/9701013].
%%CITATION = HEP-EX 9701013;%%
\bibitem{VENUS97}H. Hamasaki et al. [VENUS Collaboration],
 Phys. Lett {\bf B 407} (1997)
185.
%%CITATION = PHLTA,B407,185;%%
\bibitem{FZOZ89} G.~R.~Farrar, H.~Zhang, A.~A.~Ogloblin and I.~R.~Zhitnitsky, 
Nucl. Phys. {\bf B 311} (1989) 585.
%%CITATION = NUPHA,B311,585;%%
\bibitem{DKV03} M.~Diehl, P.~Kroll and C.~Vogt, 
Eur. Phys. J. {\bf C 26} (2003) 567 [arXiv:hep-ph/0206288].
%%CITATION = HEP-PH 0206288;%%
\bibitem{FRSW03} A.~Freund, A.~V.~Radyushkin, A.~Sch\"afer and C.~Weiss, 
Phys. Rev. Lett. {\bf 90} (2003) 
092001 [arXiv:hep-ph/0208061].
%%CITATION = HEP-PH 0208061;%%
\bibitem{GSI01}H.~H. Gutbrod et al., {\it An International 
Accelerator Facility for Beams of Ions and Antiprotons}, Conceptual 
Design Report, GSI Darmstadt (2001).
\bibitem{BS04}C.~F. Berger and W. Schweiger, in preparation.
\end{thebibliography}
\end{document}